\documentclass[aps,twocolumn,longbibliography]{revtex4}
\usepackage[dvips]{graphics,graphicx}
\usepackage[colorlinks, linkcolor=blue, citecolor=blue, urlcolor=blue, breaklinks=true]{hyperref}
\usepackage{amsmath}
\usepackage{bm}
\usepackage{color}

\begin{document}
\title{Superradiant lasing in inhomogeneously broadened\\ ensembles with spatially varying coupling}
\author{Anna Bychek$^{1}$}
\author{Christoph Hotter$^{1}$}
\author{David Plankensteiner$^{1}$}
\author{Helmut Ritsch$^{1}$}
\affiliation{
$^1$Institut f\"ur Theoretische Physik, Universit\"at Innsbruck, Technikerstr. 21a, A-6020 Innsbruck, Austria}
\date{\today}

\begin{abstract}
Theoretical studies of superradiant lasing on optical clock transitions predict a superb frequency accuracy and precision closely tied to the bare atomic linewidth. Such a superradiant laser is also robust against cavity fluctuations when the spectral width of the lasing mode is much larger than that of the atomic medium. Recent predictions suggest that this unique feature persists even for a hot and thus strongly broadened ensemble, provided the effective atom number is large enough. Here we use a second-order cumulant expansion approach to study the power, linewidth and lineshifts of such a superradiant laser as a function of the inhomogeneous width of the ensemble including variations of the spatial atom-field coupling within the resonator. We present conditions on the atom numbers, the pump and coupling strengths required to reach the buildup of collective atomic coherence as well as scaling and limitations for the achievable laser linewidth.
\end{abstract}
\maketitle

\section{Introduction}

Collective stimulated emission of coherent light by atoms inside an optical cavity is a fundamental phenomenon studied for decades in quantum optics \cite{Dicke54,bonifacio1971quantum,haake1993superradiant,benedict1996super,scully2009super,Chen09,Holland09}. Even very recently a large number of theoretical and experimental studies focused on continuous superradiance \cite{Chen09,Holland09,meiser2010steady,Maier14,
zhang2018monte,debnath2018lasing,
Hotter19,gogyan2020characterisation, Shankar21, Wu2021}, aiming at the development of a superradiant laser \cite{Thompson12, Thompson16, norcia2016superradiance, Schreck19, laske2019pulse, Schaffer20, tang2021cavity}. Such a superradiant laser typically operates in a bad-cavity regime, where the cavity mode is much broader than the natural linewidth of the atoms providing the gain. In the limit of low photon number operation the coherence necessary for frequency stability is stored in the atoms rather than the cavity field. This makes the laser frequency insensitive to thermal and mechanical fluctuations of the cavity, which is the main limitation for conventional good-cavity lasers \cite{numata2004thermal, notcutt2006contribution}. In recent years pulsed superradiance has been experimentally demonstrated \cite{Thompson12, laske2019pulse, Schaffer20, tang2021cavity} and a number of new theoretical ideas have been proposed \cite{Liu20, jager2021regular, Norcia19}. However, the experimental realization of a continuous wave superradiant laser has not yet been achieved.

Effects such as frequency broadening in the gain medium are an inherent part of any experiment. Such processes are capable of disrupting the collective interaction between the atoms and the cavity field. In this work, we aim to offer a comprehensive study of these potentially detrimental effects. To this end, we study a model of a superradiant laser and focus on inhomogeneity among the atomic ensemble. The inhomogeneity is primarily associated with a distribution of the atomic resonance frequencies leading to stimulated emission into the cavity at a range of different frequencies. Similar differences in the atom-field coupling due to variation in the atomic positioning are also included in the system.

We numerically investigate the dynamics of an atomic medium with a wide range of resonance frequencies and show how the intensity of the pumping rate can lead to cooperative effects among the atoms such that superradiant lasing is achieved. Furthermore, we consider atoms to have different coupling strengths to the cavity. We also study the laser sensitivity to cavity noise.

\section{Model}

\begin{figure}[!t]
    \centering
	\includegraphics[width=1\columnwidth]{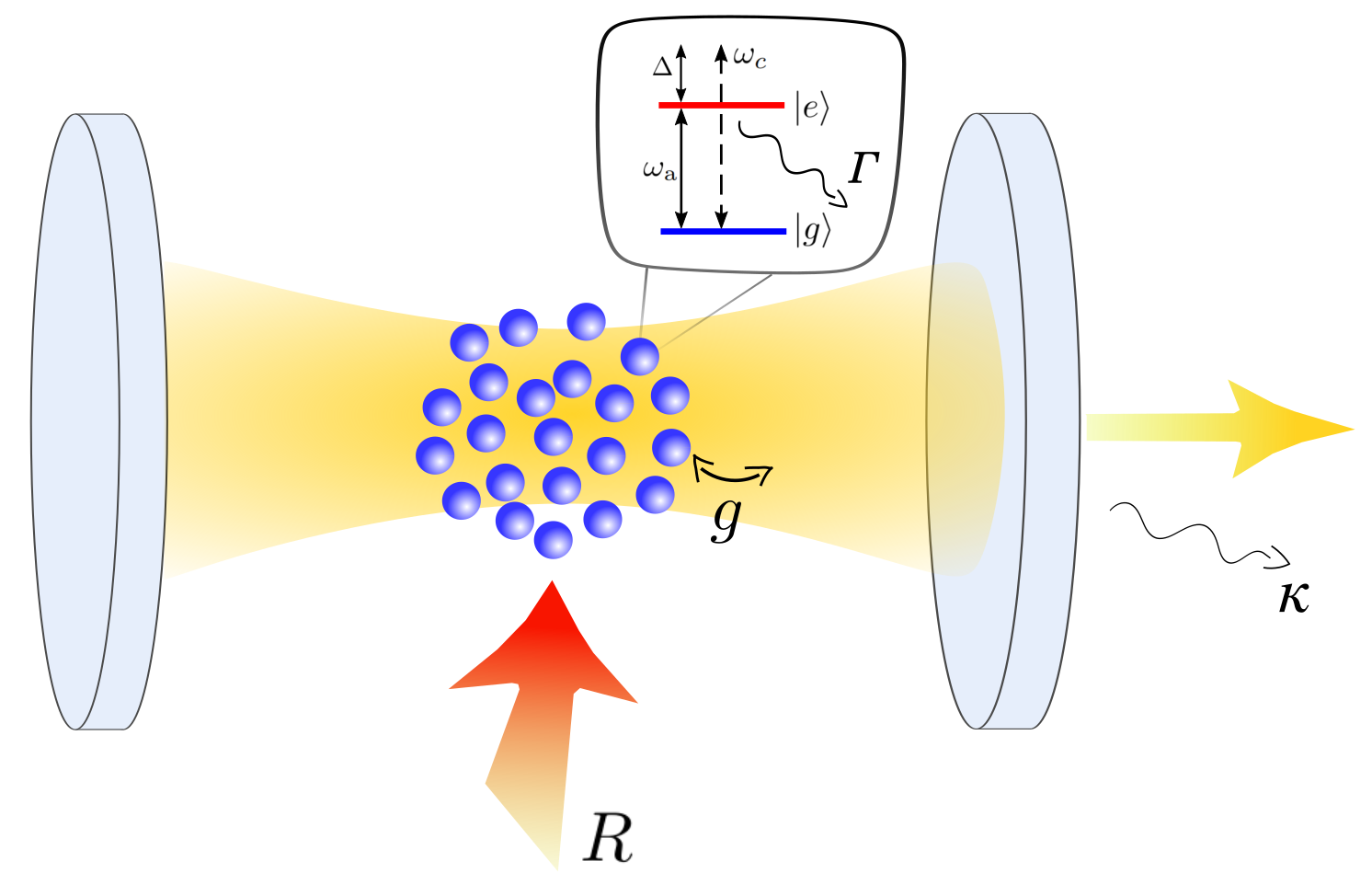}
	\caption{Schematic illustration of the system. The atomic medium is placed inside the optical resonator which has a resonance frequency $\omega_c$. Each atom features a ground and an excited state separated by the transition energy $\omega_a$. The transition couples to the cavity mode ($g$) as well as the environment ($\Gamma$). Additionally, the atoms are incoherently driven from the side ($R$) such that they can provide gain to the cavity mode.}
	\label{fig0}
\end{figure}

We consider an ensemble of $N$ incoherently pumped two-level atoms inside a single mode optical cavity as shown in Figure \ref{fig0}. In a bad-cavity regime, where the cavity relaxation rate exceeds the natural linewidth of the atomic transition by many orders of magnitude ($\kappa \gg \Gamma$), the system constitutes a generic model of a superradiant laser. The $i$-th atom couples to the cavity field with the coupling strength $g_i$ and has a resonance frequency $\omega_i$ which might be shifted from the unperturbed atomic transition frequency $\omega_\text{a}$. Assuming that the cavity is on resonance with the unperturbed atomic transition frequency, we describe the coherent dynamics of the system by the Tavis-Cummings Hamiltonian in the rotating frame of the cavity, 
\begin{equation}
    \label{eq:H}
    H = -\sum_{i=1}^N \Delta_i \sigma^+_i \sigma^-_i  +  \sum_{i=1}^N g_i (a \sigma^+_i + a^\dagger \sigma^-_i).
\end{equation}
Here, $\Delta_i = \omega_c - \omega_i$, $\sigma^+_i = (\sigma^-_i)^\dagger = | e \rangle_i \langle g |_i$ denote the raising and lowering operators of the $i$-th atom, where $| g \rangle$ and $| e \rangle $  are  the  atomic  ground  and  excited  states, respectively, and $a^\dagger$ ($a$) is the photon creation (annihilation) operator of the cavity mode. The dissipative processes of this system are described by the Liouvillian terms
\begin{equation}
\begin{aligned}
\mathcal{L}_\kappa[\rho] &= \frac{\kappa}{2}(2a\rho a^\dagger - a^\dagger a \rho - \rho a^\dagger a)\\
\mathcal{L}_\Gamma[\rho] &= \frac{\Gamma}{2} \sum_i (2\sigma^-_i \rho \sigma^+_i -\sigma^+_i \sigma^-_i \rho -\rho \sigma^+_i \sigma^-_i)\\
\mathcal{L}_R [\rho] &= \frac{R}{2} \sum_i (2\sigma^+_i \rho \sigma^-_i -\sigma^-_i \sigma^+_i \rho -\rho \sigma^-_i \sigma^+_i),\\
\end{aligned}
\end{equation} 
representing the loss of photons through the cavity at the rate $\kappa$, the spontaneous atomic decay  with the single-atom spontaneous emission rate $\Gamma$, and the individual incoherent pumping with the pump strength $R$. Thus, the full dynamics of the system is determined by the master equation for the density matrix $\rho$ in standard Lindblad form 
\begin{equation}
    \label{master equation}
\dot{\rho} = -i[H, \rho] + \mathcal{L}_\kappa[\rho] + \mathcal{L}_\Gamma[\rho] + \mathcal{L}_R[\rho].
\end{equation}
Since the exponential growth of the Hilbert space with the number of atoms renders the solution of the master equation~\ref{master equation} intractable for $N \gg 1$, we use a cumulant expansion method \cite{Holland09, Kubo62}. First, we write down the equations for operator averages describing our system, which for a given operator $\cal{O}$ reads
\begin{equation}
\label{Heisenberg0}
\begin{aligned}
\frac{d}{dt} \langle {\mathcal{O}} \rangle &= i \langle [H, {\cal{O}}] \rangle + \kappa\langle\mathcal{D}[a]\mathcal{O}\rangle +
\\
&\Gamma\sum_i\langle\mathcal{D}[\sigma_i^-]\mathcal{O}\rangle + R\sum_i\langle\mathcal{D}[\sigma_i^+]\mathcal{O}\rangle,
\end{aligned}
\end{equation}
where $\mathcal{D}[c]\mathcal{O} = \left(2c^\dagger\mathcal{O}c - c^\dagger c\mathcal{O} - \mathcal{O}c^\dagger c\right)/2$. 
We note that in some cases (mentioned in the description of the results) we additionally include cavity dephasing and atomic dephasing described by the terms $\xi \langle \mathcal{D}[a^\dagger a] \mathcal{O} \rangle$ and $\nu \sum_i \langle \mathcal{D}[\sigma_i^+\sigma_i^-] \mathcal{O} \rangle$, respectively. The cavity dephasing accounts for the effective noise imposed on the system by thermal fluctuations of the cavity mirrors, whereas the atomic dephasing models perturbations on the lasing transition.

To obtain a closed set of differential equations we use the cumulant expansion method \cite{Kubo62} up to second order:

\begin{widetext}
\begin{equation}
\label{HeisenbergN}
\begin{aligned}
&\frac{d}{dt} \langle a^\dagger a \rangle = -\kappa \langle a^\dagger a \rangle +i\sum_{m=1}^N g_m \langle a \sigma^+_m \rangle -i\sum_{m=1}^N g_m \langle a^\dagger \sigma^-_m \rangle\\
&\frac{d}{dt} \langle a \sigma^+_m \rangle = -\big( (\kappa+ \Gamma + R +\xi +\nu)/2 +i\Delta_m \big) \langle a \sigma^+_m \rangle +ig_m\langle a^\dagger a \rangle -2ig_m\langle a^\dagger a \rangle \langle \sigma^+_m \sigma^-_m \rangle -ig_m \langle \sigma^+_m \sigma^-_m \rangle -i \sum_{j; m\neq j}^N g_j \langle \sigma^+_m \sigma^-_j \rangle\\
&\frac{d}{dt} \langle \sigma^+_m \sigma^-_m \rangle = ig_m \langle a^\dagger \sigma^-_m \rangle -ig_m \langle a \sigma^+_m \rangle -(\Gamma+R) \langle \sigma^+_m \sigma^-_m \rangle +R\\
&\frac{d}{dt} \langle \sigma^+_m \sigma^-_j \rangle = ig_m \langle a^\dagger \sigma^-_j \rangle -ig_j \langle a \sigma^+_m \rangle -2ig_m\langle a^\dagger \sigma^-_j \rangle \langle \sigma^+_m \sigma^-_m \rangle +2ig_j \langle a \sigma^+_m \rangle \langle \sigma^+_j \sigma^-_j \rangle -(\Gamma +R+\nu)\langle \sigma^+_m \sigma^-_j \rangle.\\
\end{aligned}
\end{equation}
\end{widetext}
\begin{figure*}[!t]
    \centering
	\includegraphics[width=0.49\textwidth]{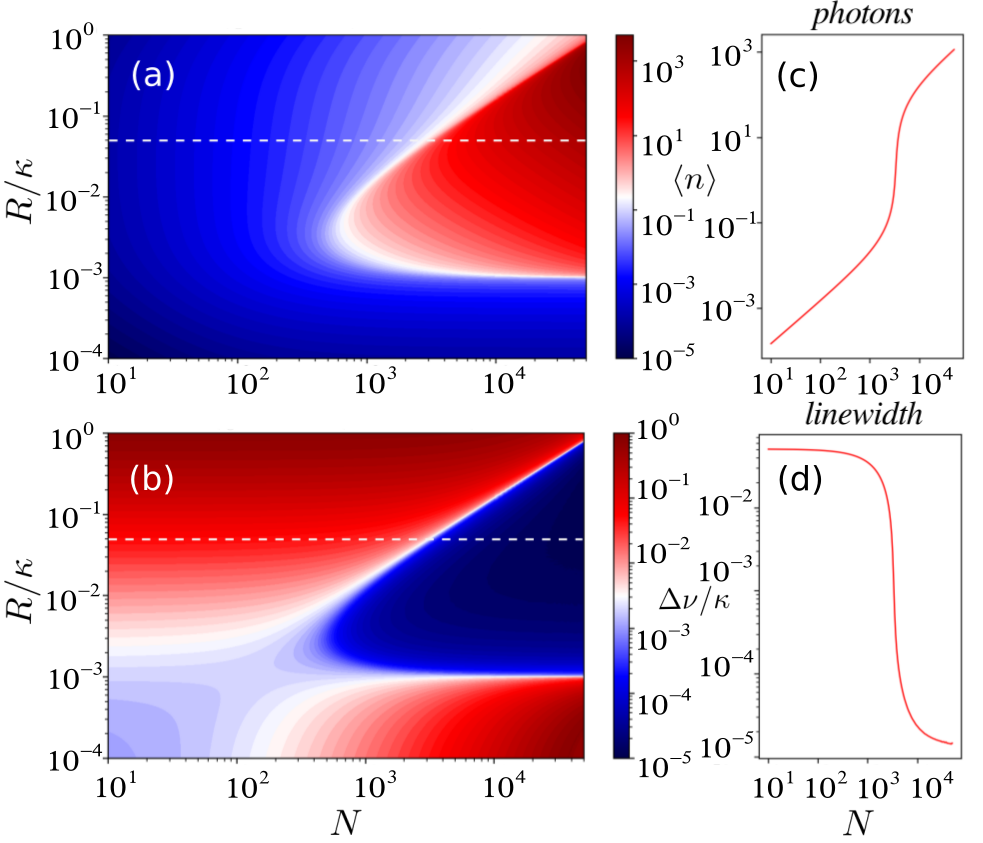}
	\includegraphics[width=0.49\textwidth]{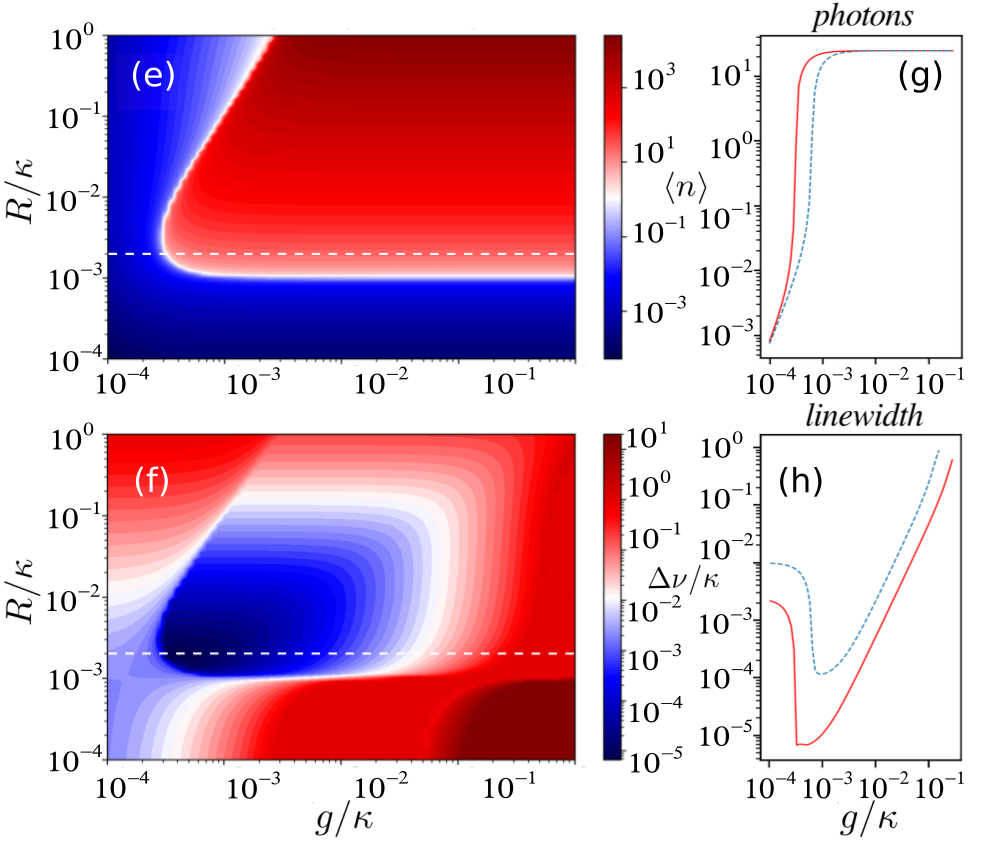}
	\caption{(a) The mean photon number and (b) the linewidth (in units of $\kappa$) as functions of the number of atoms $N$ and pumping rate $R$ for the parameter set ${(\Delta, g, \Gamma, \xi, \nu) = (0, 0.002\kappa, 0.001\kappa, 0, 0})$. (c-d) The cut through the white dashed line in (a-b) for $R=0.05\kappa$. (e-f) The mean photon number and the linewidth as functions of the atom-cavity coupling strength $g$ and pumping rate $R$. Additional cavity dephasing occurs at the rate $\xi = \kappa$. Parameters: ${\Delta=0, \Gamma=0.001\kappa, N=5\times 10^4}$. (g-h) The cut through the white dashed line in (e-f): the ultra-narrow linewidth is robust to cavity dephasing $\xi = \kappa$ (red solid line) in the regime where the photon number is low. For the blue dashed line atomic dephasing was added to the system with the rate $\nu = 10\Gamma$.}
	\label{fig1_1}
\end{figure*}
In order to calculate the spectrum of the cavity light field we make use of the Wiener–Khinchin theorem \cite{Puri01}, which states that the spectrum can be computed as the Fourier transform of the first-order correlation function $g^{(1)} (\tau) = \left\langle a^\dagger (\tau) a(0) \right\rangle $,
\begin{equation}
\label{WK_theorem}
S(\omega) = 2 \Re \left\{\int_{0}^{\infty} d\tau e^{-i\omega \tau} g^{(1)} (\tau) \right\}.
\end{equation} 
We use the quantum regression theorem \cite{Carmichael13} to write down the set of differential equations for the two-time correlation function, which in matrix form reads,
\begin{equation}
\label{Spectrum}
{\small
\frac{d}{d\tau} \begin{pmatrix} \langle a^\dagger (\tau) a(0) \rangle \\ \langle \sigma^+_1 (\tau) a(0) \rangle\\ \vdots \\ \langle \sigma^+_N (\tau) a(0) \rangle \end{pmatrix} =
\mathbf{A}
\begin{pmatrix} \langle a^\dagger(\tau) a(0) \rangle \\\langle \sigma^+_1(\tau) a(0) \rangle \\ \vdots \\ \langle \sigma^+_N(\tau) a(0) \rangle \end{pmatrix},
}
\end{equation}
where
\begin{equation}
\label{Spectrum_matrix}
{\footnotesize
\mathbf{A} = -
\begin{pmatrix}
\frac{\kappa+\xi}{2}&-ig_1&\ldots&-ig_N\\
ig_1\langle \sigma^z_1 \rangle^{st} & \frac{\Gamma+R+\nu}{2}+i\Delta_1 &\ldots & 0\\ \vdots & \vdots& \ddots &\vdots \\ ig_N\langle \sigma^z_N \rangle & 0&\ldots &\frac{\Gamma+R+\nu}{2}+i\Delta_N 
\end{pmatrix}.
}
\end{equation}
We obtain the laser emission spectrum by taking the Laplace transform of Eq. \ref{Spectrum}, where the initial conditions are the steady-state solutions of Eqs. \ref{HeisenbergN}, for example $\langle a^\dagger (\tau=0) a(0) \rangle = \langle a^\dagger a \rangle^{st}$.

In this section, we suppose that all atoms in the ensemble are identical with the same detunings $\{\Delta_i\} = \Delta$ and couplings $\{g_i\}=g$ to the cavity mode. This reduces the problem to a set of four differential equations in Eqs. \ref{HeisenbergN}.
The mean intra-cavity photon number and the laser linewidth $\Delta \nu$ (the FWHM of the spectrum) are depicted in Figure \ref{fig1_1} as functions of the number of atoms, pumping rate, and atom-cavity coupling strength.
Superradiance is expected in the parameter regime where the single-atom cooperativity parameter $C = 4g^2/(\kappa\Gamma)< 1$, but the system is in the collective strong coupling regime \cite{Holland09}, where $C N \gg 1$.
Figures~\ref{fig1_1}(a-d) show the emergence of the superradiant regime as the number of atoms increases. Above the lasing threshold the collective emission of light with an ultra-narrow linewidth is observed. In this collective regime the phases of the atomic dipoles are synchronized via photon exchange through the cavity which leads to the buildup of a collective dipole among the atoms.

\begin{figure}[!b]
    \centering
	\includegraphics[width=\columnwidth]{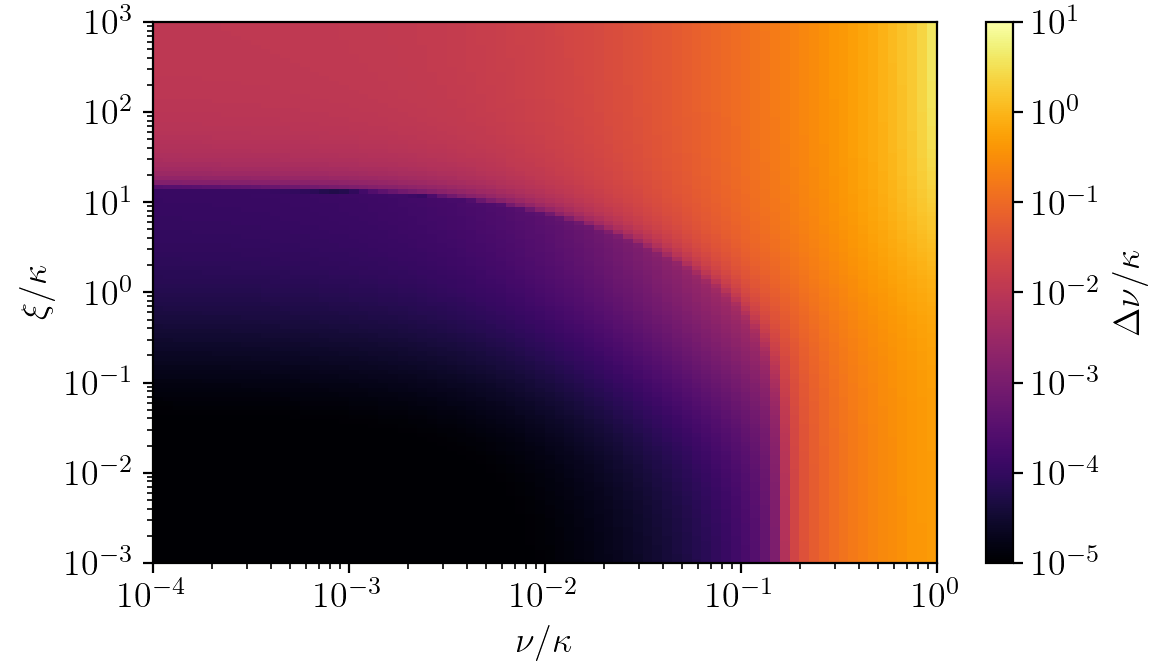}
	\caption{The linewidth of the emission spectrum of ${N=5\times 10^4}$ atoms as a scan over cavity dephasing~$(\xi)$ and atomic dephasing $(\nu)$. The optimal parameters are taken from Fig.~\ref{fig1_1}(f), where the system is in the superradiant regime for  ${(\Delta, g, \Gamma, R) = (0, 0.001\kappa, 0.001\kappa, 0.01\kappa})$.}
	\label{fig1_3}
\end{figure}

\begin{figure*}[!t]
    \centering
	\includegraphics[width=1\textwidth]{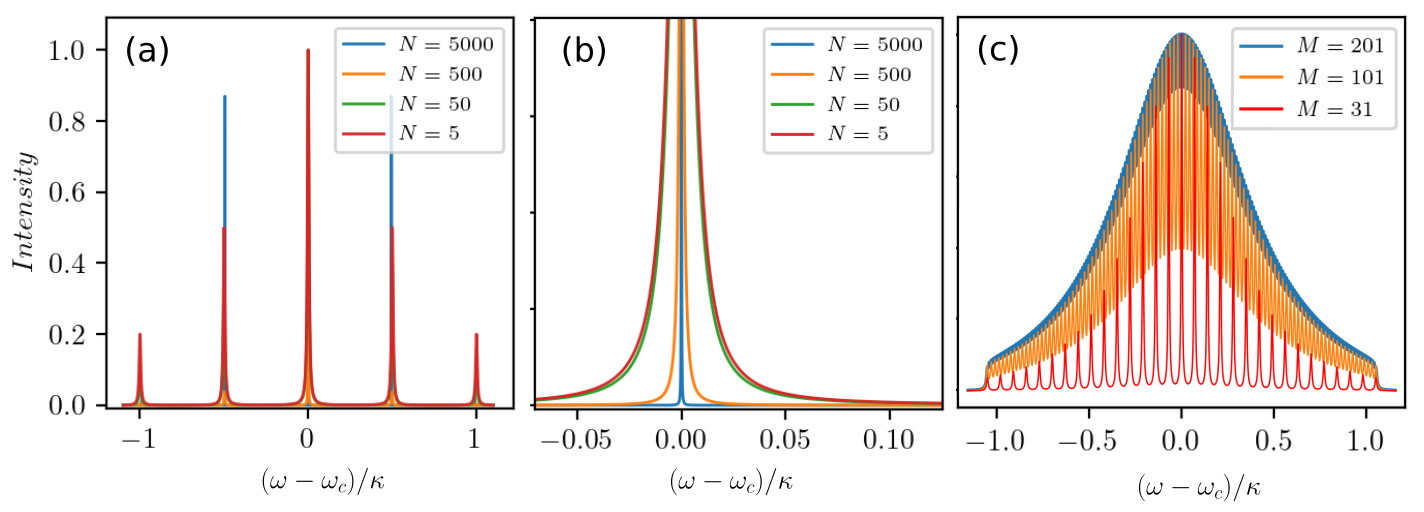}
	\caption{Cavity output spectra for weakly driven atomic ensembles composed of several discrete clusters with varying atomic frequencies. (a) $M=5$ clusters of atoms with the detunings $\Delta_m = [-\kappa; -\kappa/2; 0; \kappa/2; \kappa]$ for different total numbers of atoms $N=5,...,5000$. (b) A zoom-in showing the narrowing of the central peak in the spectrum from (a) around the resonance frequency. (c) Transition of the spectral distribution from discrete to quasi-continuous for an increasing number of clusters. Parameters: ${(g, \Gamma, R) = (0.002\kappa, 0.001\kappa, 0.01\kappa})$.}
	\label{fig2}
\end{figure*}

A key feature of such a laser is its insensitivity to thermal and mechanical fluctuations of the cavity length, since the coherence is primarily stored in the atoms rather than in the cavity field.
To show the robustness against cavity noise we include cavity dephasing with the rate $\xi$ in the equations. 
In Figure \ref{fig1_1}(f) we scan the linewidth over the coupling strength $g$ and pumping rate $R$ for an ensemble of $N=5\times 10^4$ atoms. 
In the superradiant regime, the laser linewidth is less than the natural linewidth of the atomic transition and approaches the value ${\Delta \nu \sim C \Gamma}$, which can be well below 1~mHz for the ${^1S_0} \rightarrow {^3P_0}$ transition in $^{87}\mathrm{Sr}$, as has been pointed out in Ref. \cite{Holland09}. Furthermore, we study the influence of noise on the laser linewidth in more detail. In Figure~\ref{fig1_3} we scan the linewidth over both cavity and atomic dephasing, where the other parameters of the system correspond to the superradiant regime. One can see that the linewidth of the superradiant laser can be extremely robust to noise sources within a wide range.

So far the results are based on the idea of absolutely identical atoms. In the next sections, we focus on inhomogeneity within the atomic medium. In particular, we will consider the atoms to be subject to distinct frequency shifts and different couplings to the resonator mode.

\section{Atomic ensembles with inhomogeneous broadening}

While the individual atoms in free space are identical and have the same transition frequencies in principle, in practise they are often subject to individual perturbations introducing local lineshifts, e.g. from trapping within the cavity, motion, or optical pumping. Specifically, it can be an inhomogeneous trapping lattice or pump lasers with a Gaussian profile. Doppler shifts would have similar broadening effects in ring cavities, whereas in a standing-wave cavity they would generate a time-dependent atom-field coupling which we do not consider here. In this section we study the overall effects of inhomogeneous broadening of the gain medium on the laser properties.

In contrast to the case of identical atoms, where the atom number in Eqs. \ref{HeisenbergN} and \ref{Spectrum} only enters as a constant factor, the inhomogeneity among atomic frequencies requires keeping track of the time evolution of each atom separately. For the solution of the collective dynamics one then needs to solve $\mathcal{O}(N^2)$ equations. This is only possible for a limited atom number and we thus have to resort to further approximation methods in order to treat larger ensembles. As a possible approach to approximate a large ensemble with a continuous frequency distribution we combine several atoms in subgroups representing their average atomic frequencies, which we call clusters, see also Refs. \cite{Shankar21, Wu2021, Debnath2019}. Each atom in a cluster is assumed to be completely identical to all other atoms in the same frequency cluster. This preserves the central physics of the inhomogeneous broadening, but at the same time substantially reduces the number of equations.

First, we simulate $N=5$ atoms in five clusters centered at ${\Delta_m = \omega_c - \omega_m}$, where $\Delta_m \in  [-\kappa:\kappa]$. Note that this is equivalent to $M=5$ frequency clusters each containing a single atom. At low excitation the resulting cavity output spectrum then consists of precisely five spectral lines at the frequency of each cluster.
Basically, these are five independent lasers using the same cavity mode simultaneously. 
If we increase $N$ and set the number of atoms per cluster according to a Gaussian normal distribution with the standard deviation $\sigma = \kappa$, the structure of the spectrum in Figure~\ref{fig2}(a) will remain unchanged, with each peak becoming more pronounced.
In particular, in Figure \ref{fig2}(b) we observe growing collective emission among atoms of the same cluster so that the linewidth of each peak becomes smaller as the atom number in the corresponding cluster increases. In Figure \ref{fig2}(c) we show how more and more lines appear as we increase the number of clusters up to $M=201$ until the output merges into a single broad emission line.
Note that an increase of the collective coupling to $g\sqrt{N} \sim \kappa$ or a randomization of the individual cluster detunings do not lead to any substantial difference in the spectral profile of the laser. Hence, one can expect a single broadened peak in the emission spectrum in the more realistic case of a large ensemble of atoms with a continuous frequency distribution.

\begin{figure}[!t]
    \centering     
	\includegraphics[width=0.5\textwidth]{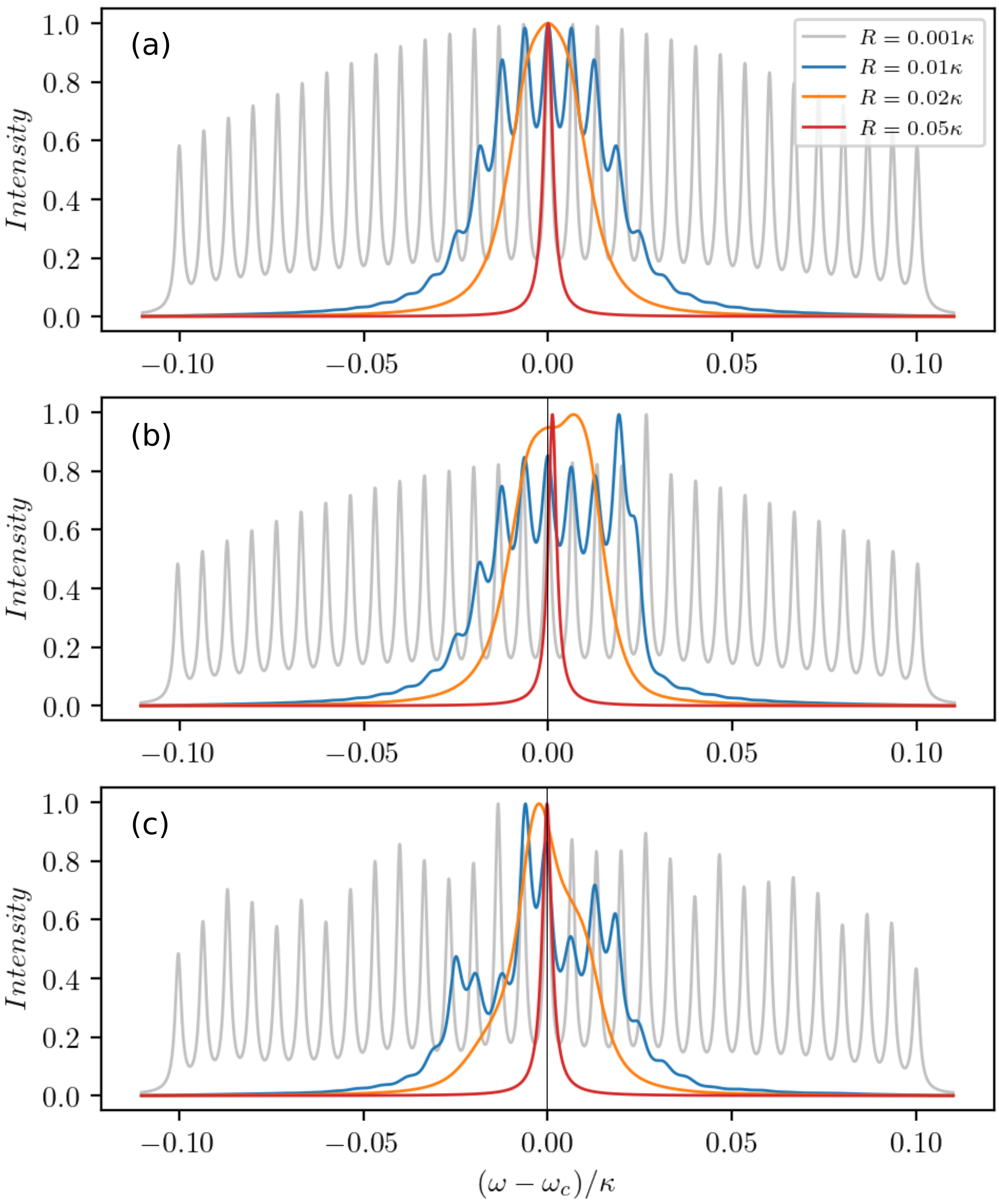}
	\caption{Cavity output spectra of a large inhomogeneously broadened ensemble of $N=10^4$ atoms for different pumping rates $R=0.001\kappa$ (grey), $0.01\kappa$ (blue), $0.02\kappa$ (orange), $0.05\kappa$ (red). The ensemble is represented by $M=31$ clusters with the number of atoms per cluster chosen according to a Gaussian normal distribution (a) with the standard deviation $\sigma=0.1\kappa$, (b) when adding particle imbalance at $\Delta = 0.027\kappa$, (c) with overall atom number fluctuations. The emission intensity is normalized and the other parameters are chosen as ${\Delta \in [-\sigma:\sigma]}$, $g=0.002\kappa$, $\Gamma=0.001\kappa$. }
	\label{fig3}
\end{figure}

So far we limited investigations to weak incoherent pumping in order to avoid significant additional broadening of the atomic linewidth due to pumping. However, this broadening effect can actually aid the buildup of coherences between the clusters. When the pumping is strong enough such that the distinct spectral lines overlap, the discrete spectral lines of the clusters merge into a single central peak (see Figure~\ref{fig3}). In other words, more intra-cavity photons and broader individual atomic gain lines ultimately lead to a dramatic narrowing of the laser line.  We attribute this effect to a dynamical phase transition from the unsynchronized phase of the dipoles to the synchronized one. 
Note that an analogous phenomenon has previously been studied in Ref. \cite{Xu14} for two mesoscopic ensembles of atoms collectively coupled to a cavity with opposite detunings.
Furthermore, we show how an atom number imbalance at a particular frequency in Figure~\ref{fig3}(b) and overall atom number fluctuations modeled by slight random deviations from a Gaussian distribution in Figure~\ref{fig3}(c) lead to a shift of the spectral lines. However, in the synchronized regime the lineshift of the central peak is much smaller than its linewidth.

\begin{figure}[!t]
    \centering
	\includegraphics[width=0.5\textwidth, height = 3.72cm]{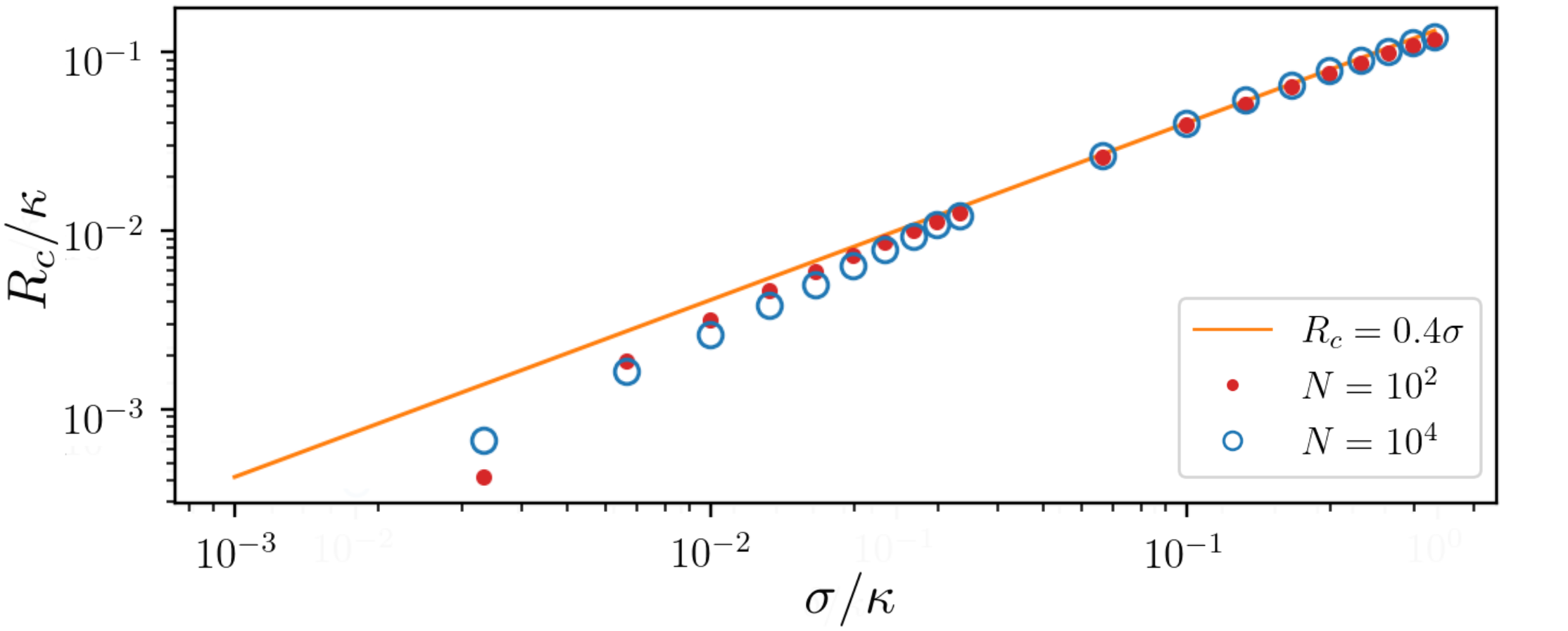}
	\caption{Critical value of pumping above which the collective superradiant regime is established depending on the standard deviation $\sigma$ of the atomic frequency distribution. The data points show the numerical results for an ensemble of ${N=10^2}$ (red dots) and ${N=10^4}$ (blue circles) atoms sampled by ${M=31}$ clusters. For comparison we plot the linear (solid line) function $R_c=0.4\sigma$ as a linear approximation to the data points. Parameters: ${\Delta \in [-3\sigma:3\sigma], g=0.001\kappa, \Gamma=0.001\kappa}$.}
	\label{fig3b}
\end{figure}

The collapse of the emission spectrum into a single central line occurs at a critical pump strength $R_c$. This critical value strongly depends on the overall width of the frequency distribution, but shows almost no dependence on the number of subensembles $M$ and the total number of atoms $N$.
The critical transition pump strength is shown for different standard deviations $\sigma$ of the atomic frequency distribution in Figure \ref{fig3b}.
The data points show the numerical results for an ensemble of $N=10^2$ (red dots) and $N=10^4$ (blue circles) atoms sampled by $M=31$ clusters. For comparison, we also plot the linear (solid line) function $R_c=0.4\sigma$. We calculate the critical pumping by computing the spectrum for different $R$. We then determine the critical value of the pump strength as the value at which the spectrum has only a single local maximum, i.e. all separate peaks have merged into a single spectral line. We find a linear dependence for large inhomogeneously broadened ensembles while for narrow ensembles a significantly lower pump strength is required.

\begin{figure}[!t]
    \centering
	\includegraphics[width=\columnwidth]{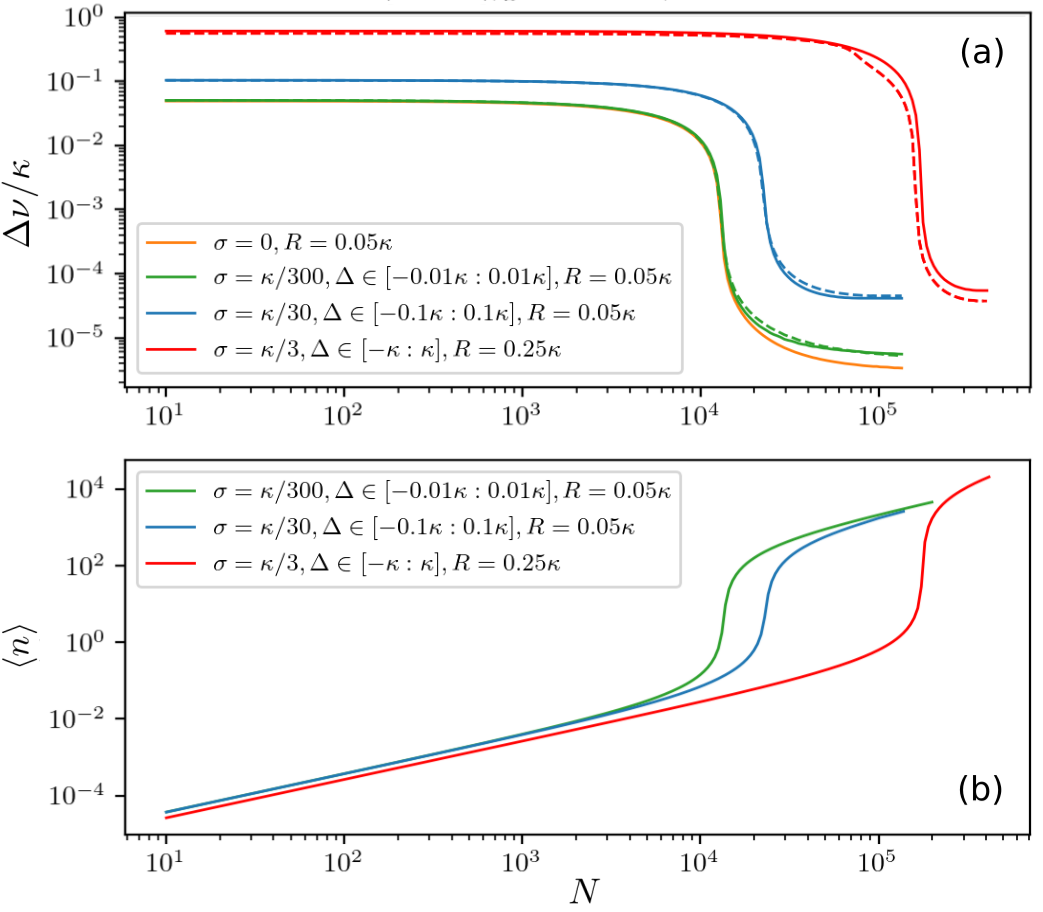}
	\caption{(a) Laser spectral linewidth and (b) mean photon number for inhomogeneously broadened ensembles with different standard deviations $\sigma$ and spectral widths of atomic frequencies ${\Delta \in [-3\sigma:3\sigma]}$, where $\sigma=\kappa/300$ (green line), $\sigma=\kappa/30$ (blue line), $\sigma=\kappa/3$ (red line) as a function of the total number of atoms. The number of clusters is $M=31$ with the number of atoms per cluster chosen according to a Gaussian normal distribution. 
	The dashed lines represent the results including an additional spatial variation of the atom-field coupling $g(x) = g_0 \cos(kx)$. The ensemble is comprised of $M = 11$ frequency clusters and $K=5$ clusters of different couplings. The couplings are chosen such that the effective coupling strength $g_\mathrm{eff} = \sqrt{(\sum_m g_m^2)/K} \equiv g$. Parameters: ${g=0.001\kappa,~g_0=0.0013\kappa,~\Gamma=0.001\kappa}$. }
	\label{fig4}
\end{figure}

Once the laser is operating at a single distinct emission frequency, we can characterize the properties of the output light by the linewidth and the average photon number. 
The results for different distributions of atomic frequencies are shown in Figure \ref{fig4}, where ${\Delta \in [-3\sigma:3\sigma]}$ and ${\Gamma \leq 3\sigma \leq \kappa}$.
Figure~\ref{fig4}(a) illustrates how a narrow linewidth appears for different $\sigma$ as the number of atoms increases. Note that we chose a pumping strength well above the critical value for a wide atomic frequency distribution (red line). The sharp decrease of the linewidth is accompanied by an increase in the average photon number as can be seen in Figure \ref{fig4}(b). This is indicative of a lasing threshold being crossed at a certain number of atoms.

\section{Inhomogeneously broadened ensembles with variable coupling strength}

\begin{figure}[!t]
    \centering
	\includegraphics[width=\columnwidth]{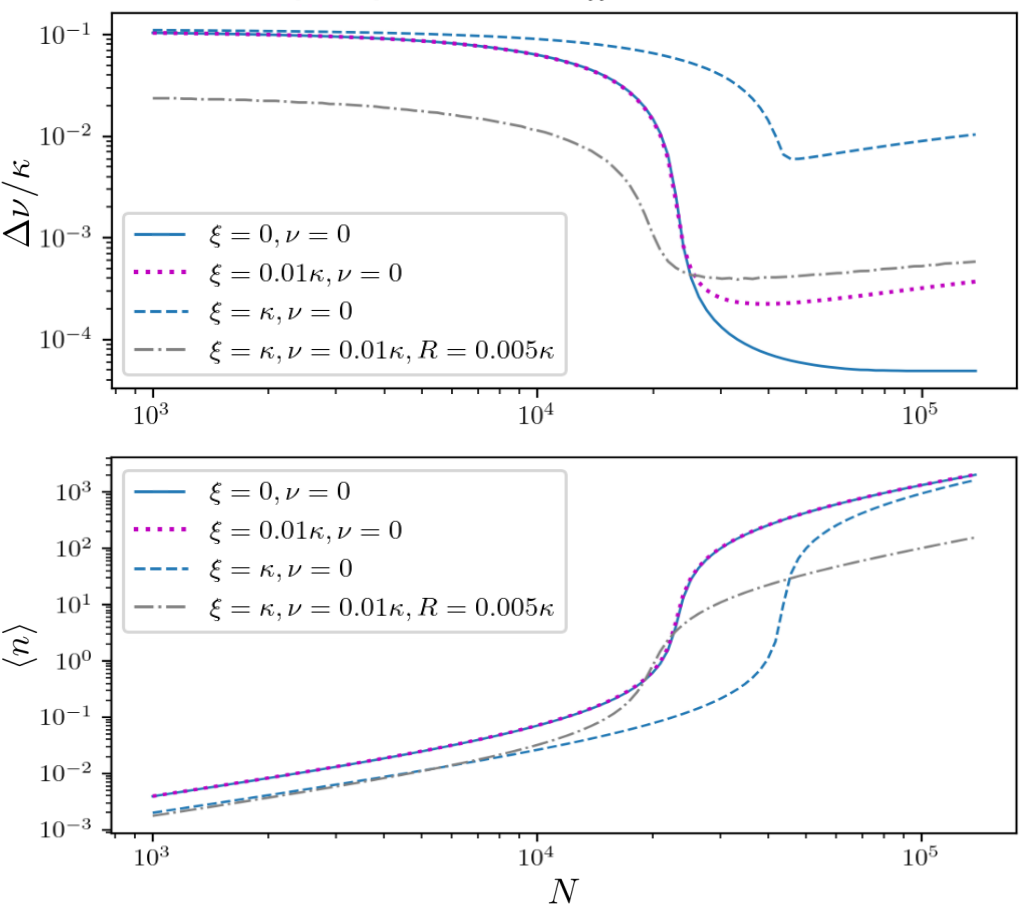}
	\caption{Laser linewidth (upper panel) and mean photon number (lower panel) for an inhomogeneously broadened ensemble with spatially varying coupling for ${\sigma=\kappa/30}$, ${\Delta \in [-0.1\kappa:0.1\kappa]}$ and $R=0.05\kappa$ (solid line). Adding various cavity dephasing at the rate $\xi=\kappa$ (dashed blue line) and $\xi=0.01\kappa$ (dotted magenta line) we can identify an optimal atom number, above which the cavity noise overwhelms the linewidth narrowing due to large photon numbers. The dash-dotted grey line shows the results when adding additional atomic dephasing at the rate $\nu=0.01\kappa$. This additional broadening allows synchronization of the individual clusters in the weak pumping regime $R=0.005\kappa$ ultimately leading to a smaller linewidth.
	}
	\label{fig5}
\end{figure}

Up to now we have assumed that the atoms are perfectly positioned inside the cavity such that they couple equally to the cavity mode. Let us now include spatial variations of the atom-field coupling within the resonator. We consider the ensemble of atoms with the position-dependent coupling strength $g(x) = g_0 \cos(kx)$, where $g_0$ is the coupling constant, $k=2\pi/\lambda$ is the cavity mode wave number and $x$ represents the position of an atom.
In order to describe the atom-field dynamics we use a similar cluster approach as before. We assume equidistant positions for different clusters $x_m \in [0,..., \lambda/4 )$ and corresponding couplings $g_m(x) = g_0 \cos(kx_m) = \{g_1, g_2, ..., g_K \}$, where $K$ is the total number of clusters. Note, that the sign of the coupling is irrelevant in our system, therefore we only consider couplings with $g_m > 0$.

The dashed lines in Figure \ref{fig4} show the results for ${M=11}$ frequency clusters and $K=5$ clusters of different couplings. As can be seen in Figure \ref{fig4}(a), for atoms with different couplings to the cavity mode the dependence of the linewidth on the number of atoms remains roughly the same as for atoms equally coupled to the cavity. This holds as long as the effective overall coupling strength $g_\mathrm{eff} = \sqrt{(\sum g_m^2)/K}$ is constant. Thus, the linewidth is essentially unaffected by atoms having different couplings to the cavity.

Finally, let us include cavity dephasing in order to describe lasing in a large inhomogeneously broadened ensemble in the presence of cavity noise. The spectral linewidth and mean photon number under strong cavity dephasing at the rate $\xi=\kappa$ are depicted in Figure \ref{fig5} (blue dashed line). Note that establishing coherence in such a largely broadened ensemble requires sufficiently strong pumping. This subsequently leads to a large number of photons in the cavity mode making the setup sensitive to cavity fluctuations, see Figure \ref{fig1_1}(f). However, additional atomic dephasing can actually relax the constraint on the pumping, since both incoherent pumping and atomic dephasing are closely tied to the same physical effect of broadening the atomic emission line. 
Thus individual atomic dephasing induce additional atom-atom coupling by enlarging the overlap of distinct spectral lines, which finally leads to better synchronization. Adding atomic dephasing to the system at the rate $\nu=0.01\kappa$ allows for maintaining collective interactions in the ensemble and at the same time enables a reduction of the pump strength by one order of magnitude to $R=0.005\kappa$.
In the low photon number regime, a linewidth on the order of the natural atomic linewidth $\Gamma$ can be achieved in the presence of strong atomic and cavity dephasing (dash-dotted grey line).

\section{Conclusions}

We studied superradiant lasing when the gain medium is subject to substantial inhomogeneous frequency broadening and variable coupling. In extensive numerical simulations based on a second-order cumulant expansion we were able to confirm previous predictions that sufficiently large numbers of atoms subject to strong optical pumping can induce synchronization of the atomic dipoles over a large bandwidth. This generates collective stimulated emission of light into the cavity mode leading to narrow-band laser emission at the average of the atomic frequency distribution. The linewidth is orders of magnitudes smaller than that of the cavity as well as the inhomogeneous gain broadening and exhibits reduced sensitivity to cavity frequency noise. We determine the operational conditions and, in particular, the best pump rate to choose for achieving the smallest linewidth for a given atom number and cavity. The minimum occurs not at very low photon numbers but at intra-cavity photon numbers reaching a significant fraction of the atom number.

Typically, full synchronization requires fairly strong pumping, which increases the effective atomic linewidth. We determined the minimum pump strength to achieve collective phase-locked oscillation of all atomic dipoles. Interestingly, some individual line-broadening effects such as atomic dephasing can actually induce synchronization at significantly lower pump rates. Furthermore, our simulations also show that variations in the atom-field coupling strength induced by the cavity mode structure play only a minor role for the laser stability and noise. In fact, they can be compensated by an increase of the effective overall coupling using a larger atom number or stronger pump.

In the present work, we did not take into account collisions or dipole-dipole interactions between atoms. The effect of dipole-dipole interactions have been studied in a small-scale full quantum model in Ref. \cite{Maier14} and do not appear too detrimental. Moreover, collisions could even have a positive effect on synchronization \cite{Zhu15} but a quantitative prediction is complicated.
So far our model is still based on a very simplistic effective pump description via an individual, independent and equal pump rate for each atom. More detailed studies of optical pumping schemes including the shifts induced by the pump light will be at the center of future studies. 


\acknowledgments

We acknowledge funding from the European Union's Horizon 2020 research and innovation program under the Marie  Sklodowska-Curie Grant Agreement No. 860579 MoSaiQC (A.~B.) and Grant Agreement No. 820404 iqClock (C.~H., D.~P., H.~R.). 

\section*{Software availability}

The presented results can be reproduced by using the source code \emph{N$\_$atoms$\_$M$\_$clusters$\_$Delta.jl}  \cite{bychek_2021_git}. The file contains an example of the cluster approach written in Julia~version~1.5.0 using the parameters in Figure \ref{fig3}. Numerical simulations were performed with the open-source framework Differentialequations.jl \cite{rackauckas2017differentialequations}. The toolbox QuantumCumulants.jl \cite{plankensteiner2021quantumcumulants} has been used to check the equations and verify the second-order cumulant expansion. The graphs were produced using the Matplotlib library~\cite{hunter2007matplotlib}.

\newpage

\begin{widetext}

\section*{Appendix. Cross-correlations between atoms in different clusters.}

\begin{figure}[!h]
    \centering
	\includegraphics[width=\columnwidth]{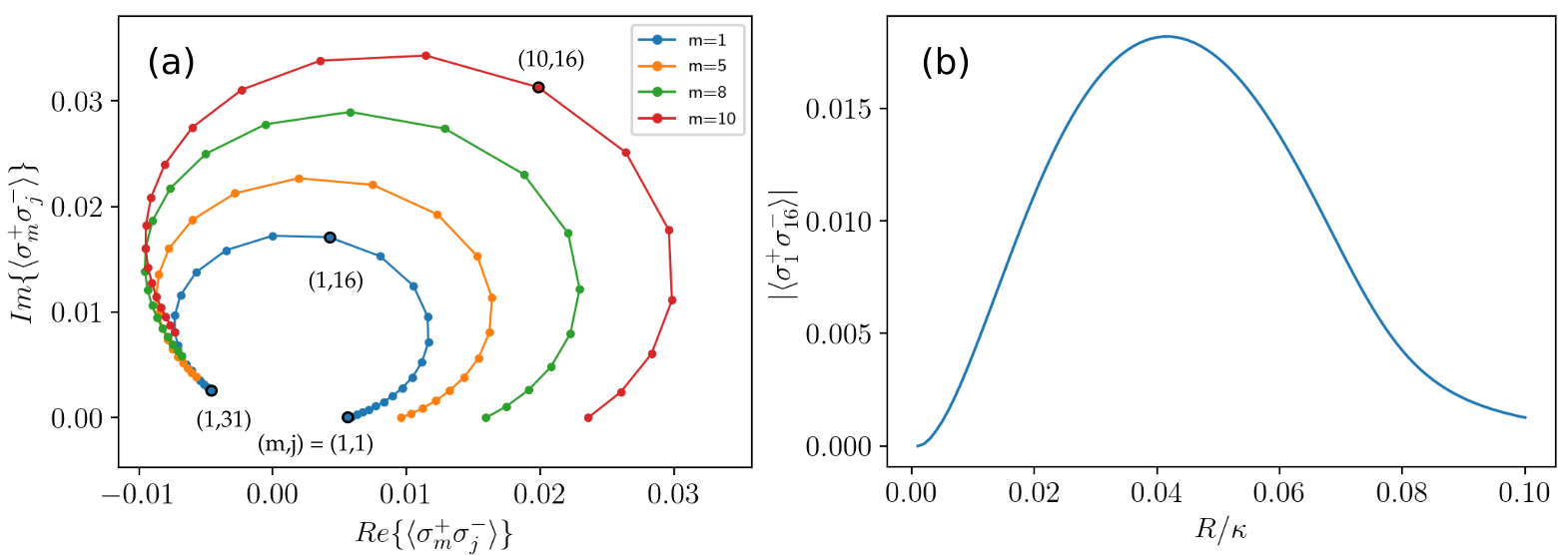}
	\caption{Cross-correlations between the 31 clusters presented in Figure \ref{fig3}(a). (a) Real and imaginary part of $\langle \sigma^+_m \sigma^-_j \rangle$ correlations between atoms in the $m$-th and $j$-th clusters on the complex plane for $R=0.05\kappa$. (b) The magnitude of the cross-correlations between atoms in the first and the central clusters as a function of the pumping strength. 
	}
	\label{fig_appendix}
\end{figure}

As we refer to in the main text, we model a continuous atomic frequency distribution with the standard deviation $\sigma$ by choosing equidistant cluster detunings $\Delta_m$ with the number of atoms per cluster $N_m$ given by a Gaussian distribution with the standard deviation $\sigma$. The Heisenberg equations for an ensemble of $N$ atoms sampled by $M$ clusters can be written as
\begin{equation}\tag{A1}
\label{HeisenbergM}
\begin{aligned}
& \frac{d}{dt} \langle a^\dagger a \rangle = -\kappa \langle a^\dagger a \rangle +\sum_{m=1}^M i g_m N_m\langle a \sigma^+_m \rangle -\sum_{m=1}^M i g_m N_m \langle a^\dagger \sigma^-_m \rangle\\
& \frac{d}{dt} \langle a \sigma^+_m \rangle = -\big( (\kappa+ \Gamma + R)/2 +i\Delta_m \big) \langle a \sigma^+_m \rangle +ig_m\langle a^\dagger a \rangle -2ig_m\langle a^\dagger a \rangle \langle \sigma^+_{am} \sigma^-_{am} \rangle -ig_m \langle \sigma^+_{am} \sigma^-_{am} \rangle + \ldots \\
& \quad \ldots -ig_m(N_m -1)\langle \sigma^+_{am} \sigma^-_{bm} \rangle -\sum_{j; m\neq j}^M i g_j N_j \langle \sigma^+_m \sigma^-_j \rangle\\
& \frac{d}{dt} \langle \sigma^+_{am} \sigma^-_{am} \rangle = ig_m \langle a^\dagger \sigma^-_m \rangle -ig_m \langle a \sigma^+_m \rangle -(\Gamma+R) \langle \sigma^+_{am} \sigma^-_{am} \rangle +R\\
& \frac{d}{dt} \langle \sigma^+_{am} \sigma^-_{bm} \rangle|_{a \neq b} = ig_m \langle a^\dagger \sigma^-_m \rangle (1-2\langle \sigma^+_{am} \sigma^-_{am} \rangle) -ig_m \langle a \sigma^+_m \rangle (1-2\langle \sigma^+_{am} \sigma^-_{am} \rangle) -(\Gamma +R)\langle \sigma^+_{am} \sigma^-_{bm} \rangle.\\
& \frac{d}{dt} \langle \sigma^+_m \sigma^-_j \rangle|_{m \neq j} = -i(\Delta_m - \Delta_j)\langle \sigma^+_m \sigma^-_j \rangle + ig_m \langle a^\dagger \sigma^-_j \rangle (1 - 2\langle \sigma^+_{am} \sigma^-_{am} \rangle) -ig_j \langle a \sigma^+_m \rangle (1 - 2\langle \sigma^+_{aj} \sigma^-_{aj} \rangle) -(\Gamma +R)\langle \sigma^+_m \sigma^-_j \rangle,\\
\end{aligned}
\end{equation}
where indices $a, b$ refer to an atom, and $m, j$ are cluster indices. The last equation describes the cross-correlations between atoms in different clusters. Next, we study the phase and the amplitude of these correlations as the system reaches the steady-state. In the weak pumping regime, the correlations are zero and therefore there is no coherence between the distinct spectral lines of the output spectra in Figure \ref{fig2}. However, in the synchronized regime shown in Figure \ref{fig3}(a) for $R=0.05\kappa$, the existing cross-correlations of the $m$-th cluster with the other clusters $j=m..M$ are presented in Figure \ref{fig_appendix}(a). 

Let us follow these correlations as the system goes from the unsynchronized phase to the synchronized one. We study the magnitude of cross-correlations between the first (outer) cluster and the central cluster in Figure \ref{fig3}(a) as a function of the pumping strength. The correlations are zero in the weak pumping regime and grow with the pumping strength as shown in Figure \ref{fig_appendix}(b).
The function reaches its maximal value when the ensemble is fully synchronized. However, as pumping continues to grow the correlations decrease due-to growing dephasing imposed by pumping. 

\end{widetext}

\newpage

\bibliography{Superradiant_laser}

\end{document}